\documentclass[%
 reprint,
nofootinbib,
 amsmath,amssymb,
 aps,
]{revtex4-2}
\usepackage[dvipsnames]{xcolor}

\definecolor{mygrey2}{RGB}{114,112,112}
\definecolor{mygrey1}{RGB}{147,145,145}
\definecolor{mygrey3}{RGB}{150,150,150}
\usepackage[normalem]{ulem}
\usepackage{soul}

\newcommand\blue{\textcolor{blue}}

\usepackage{graphicx}
\usepackage{sidecap}
\usepackage{dcolumn}
\usepackage{bm}

\usepackage[utf8]{inputenc} 
\usepackage[T1]{fontenc}    
\usepackage[hidelinks]{hyperref}       
\usepackage{url}     
\usepackage{booktabs}       
\usepackage{tabularx}
\usepackage{siunitx}
\usepackage{amsfonts}       
\usepackage{nicefrac}       
\usepackage{microtype}      
\usepackage{lipsum}		
\usepackage{tikz}
\usepackage{mathrsfs} 
\usepackage{amssymb}
\usepackage{cleveref}
\usepackage{academicons}
\usepackage{enumerate} 
\definecolor{orcidlogocol}{HTML}{A6CE39}
\usepackage{float}

\usepackage{etoolbox}
\newcommand{\appendixfigures}{%
  \renewcommand{\thefigure}{\thesection\arabic{figure}}%
  \setcounter{figure}{0}%
}

\newcounter{Snumber}

\crefname{Sequation}{Eq.~(S\theSequation)}{Eqs.~(S\theSequation)}

\newcounter{Sfigure}
\renewcommand{\theSfigure}{S\arabic{Sfigure}}

\newcounter{STable}

\crefname{Sfigure}{Fig.}{Figs.}
\crefname{SSCfigure}{Fig.}{Figs.}
\crefname{STable}{Tab.}{Tabs.}

\AtBeginDocument{%
    \crefname{equation}{Eq.}{Eqs.}%
    \crefname{chapter}{Ch.}{chapters}%
    \crefname{section}{Sect.}{Sects.}%
    \crefname{appendix}{appendix}{appendices}%
    \crefname{enumi}{item}{items}%
    \crefname{footnote}{footnote}{footnotes}%
    \crefname{figure}{Fig.}{Figs.}%
    \crefname{table}{Tab.}{Tabs.}%
    \crefname{theorem}{theorem}{theorems}%
    \crefname{lemma}{lemma}{lemmas}%
    \crefname{corollary}{corollary}{corollaries}%
    \crefname{proposition}{proposition}{propositions}%
    \crefname{definition}{definition}{definitions}%
    \crefname{result}{result}{results}%
    \crefname{example}{example}{examples}%
    \crefname{remark}{remark}{remarks}%
    \crefname{note}{note}{notes}%
}

\newcommand*\diff{\mathop{}\!\mathrm{d}}

\makeatletter
\newcommand*{\transpose}{%
  {\mathpalette\@transpose{}}%
}
\newcommand*{\@transpose}[2]{%
  \raisebox{\depth}{$\m@th#1\intercal$}%
}
\makeatother

\begin{document}

\preprint{APS/123-QED}

\title{Mechanical stress induced by the polymerization of an active gel near a surface}

\author{Kristiana Mihali}
\author{Dennis W{\"o}rthm{\"u}ller}%
\author{{Pierre Sens}}%
 \email{pierre.sens@curie.fr}
\affiliation{Institut Curie, PSL Research University, CNRS UMR 168; F-75005 Paris, France}

\date{\today}
\begin{abstract}

Actin flow in the cortical cytoskeleton underneath the cell membrane generates mechanical stresses that shape the cell surface.
We study this mechanism using a hydrodynamic model of a compressible active gel polymerizing at the membrane and undergoing turnover.
We determine how actin flow, density relaxation and friction of actin with the membrane generate stress on a corrugated membrane at the linear order in deformation. Analytical solutions in limiting regimes, combined with finite element methods in the general case, provide a map of normal and tangential stresses as functions of compressibility, interfacial friction and actin turnover, and determine the conditions under which actin polymerization can render the membrane linearly unstable. The non-linear regime is also briefly discussed.
\end{abstract}

\maketitle


Cells continually sense and reshape themselves in response to mechanical and geometrical features of their environment. A central player in this mechano-geometrical response is the actin cortex: a dynamic network of polymerizing filaments that supports the plasma membrane and generates stresses capable of deforming it \cite{salbreux2012}. A key question is how the cortical flows arising from actin polymerization can transmit normal and tangential forces to a curved membrane and how such forces depend on the properties of the cortex coupled to the geometry of the membrane. 
In a recent work, \cite{Mihali2025}, we proposed a hydrodynamic theory for a free-standing viscous and incompressible actin gel growing on a corrugated membrane. Curvature sensitive actin nucleators such as BAR domain proteins  \cite{Zhao:2011} modulate the local actin polymerization rate in a curvature-dependent manner. This feedback destabilizes a flat membrane when the curvature-sensitivity of actin polymerization exceeds a threshold. This work established how curvature-dependent normal stress emerges from growth-driven flow and may amplify geometric perturbations.

The analytical results reported in \cite{Mihali2025} relied on the simplifying assumptions that the cortex can be treated as an incompressible viscous fluid. The viscous assumption is justified for phenomena involving time scales larger than the visco-elastic relaxation time of the actin gel. However the incompressibility condition is more questionable. While it allows for analytical progress, its relevance must be assessed precisely, which is done in the present work. In the incompressible limit, we have previously shown that  force transmission between the actin gel and the membrane is independent of the friction between the two materials, which convert relative lateral movement into tangeantial (shear) stress \cite{Mihali2025}. We show here that the gel stress on the membrane does depend on friction for a compressible gel
and we quantitatively assess the importance of this
parameter in the transmission of polymerization stress and interface stability. In this work, we focus on the activity stemming from polymerization and turnover, and disregard other kinds of activity, such as the heterogeneous contractile activity generated by molecular motors (myosin) which have also been shown to lead to large fluctuation with complex dynamics \cite{Levernier:2020}.

\vspace{-0.5cm}
\section{Model}
\begin{figure}[b]
    \centering
    \vspace{-0.9cm}
    \includegraphics[width=\linewidth]{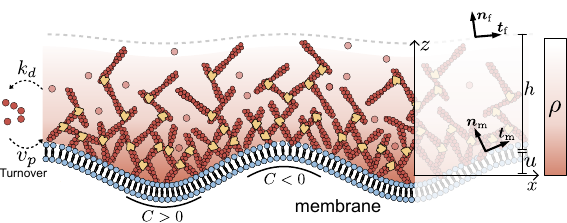}
    \caption{A viscous, compressible actin gel polymerizing on a wavy membrane. Actin polymerizes at a speed $v_p$ at the membrane and depolymerizes at a rate $k_d$ throughout the layer, resulting in an actin density $\rho$ decaying away from the membrane. This sets the layer thickness $h$.}
    \label{fig:sketch}
\end{figure}

We describe a compressible, viscous actin gel that polymerizes against a corrugated membrane described by its height $u(x)$ compared to some reference (flat) plane. This so-called Monge representation is appropriate for moderate deformations that do not have overhangs. In what follows, we mostly concentrate on small deformation (|$\nabla u|\ll 1$) for which any shape can be expanded in Fourier modes: $u(x)=\sum _q u_qe^{iqx}$. The actin network polymerizes at the membrane interface and occupies the half-space $z\geq 0$. Actin monomers are continuously added at the membrane at a rate characterised by a polymerization velocity $v_p$. Filament turnover removes polymerized actin at a constant rate $k_d$ throughout the gel layer. This situation is represented in Fig.\ref{fig:sketch}.

\vspace{-0.5cm}
\subsection{Constitutive law and mass conservation.}\label{BC}
The actin density $\rho$ and the velocity field $\boldsymbol{v}$, in the presence of turnover, satisfy the mass conservation equation
\begin{equation}
    \partial_t \rho + \nabla \cdot (\rho \boldsymbol{v}) = -k_d \rho\;.
    \label{eq:mass-conservation-compressible}
\end{equation}
We model the gel as a compressible Newtonian fluid satisfying the constitutive stress relation
\begin{equation}
    \boldsymbol{\sigma} = (-P(\rho)
   +\eta_b\nabla \cdot \boldsymbol{v}) \boldsymbol{I}+\eta (\nabla \boldsymbol{v} + \nabla\boldsymbol{v}^{\text{T}})\label{stress}\;.
\end{equation}
Here $\eta$ and $\eta_b$ are the shear and bulk viscosity, respectively, and $P(\rho)$ is the density-dependent gel pressure. For simplicity, we assume that $\eta$ is constant and disregard the contribution of bulk viscosity ($\eta_b=0$) in the main text. The contribution of $\eta_b$ to the various expressions for the stress is given in Appendix \ref{2-viscosity}.
At low Reynolds number and in the absence of body forces, the local force balance reads
\begin{equation}
    \nabla\cdot\boldsymbol{\sigma}=0\;.
\end{equation}

For an incompressible gel, the pressure $P$ acts as a Lagrange multiplier that enforces a constant gel density $\rho=\rho^*$. For a fully compressible gel, the pressure vanishes, and the density can vary without mechanical consequences. The visco-elastic properties of actin gels \textit{in-vitro} have been extensively characterised and are strongly influenced by the density and flexibility of filaments and cross-linkers and their turnover rate \cite{Xu1998,salbreux2012,Fischer-Friedrich2016}. One can generally expect a short-time elastic behaviour with long-time viscous-elastic relaxation, with a crossover influenced by the cross-linkers lifetime. However these studies on stabilised filaments do not recapitulate the effect of turnover, which is central to the present study. Actin turnover should lead to a compressible viscous fluid behaviour at long-time \cite{Julicher2007} but also controls the thickness of the gel in our case, where polymerization occurs at the membrane. In this long-time regime, actin turnover, crosslink dynamics and related microscopic processes continuously rearrange actin filaments and crosslink so the density of actin can vary. Local compression or dilution of the network will appear in a coarse-grained level as a restoring stress that will be proportional to the density deviating from the target value $\rho^*$, with proportionality coefficient $\chi$. Here we adopt a linear relationship between pressure and density 
\begin{equation}
P=\chi(\rho-\rho^*)
\end{equation}
where $\chi$ is a phenomenological gel bulk modulus. This expression does not directly include visco-elastic relaxation. Therefore the modulus $\chi$ should be considered as a phenomenological parameter that allows to continuously extrapolate between the fully compressible ($\chi=0$) and incompressible ($\chi\rightarrow\infty$) cases. For instance, both the compressibility $\chi$ and the target density $\rho^*$ would be influenced by the presence of a uniforme density of contractile units (molecular motors) that tend to increase the actin density.

\subsection{Boundary conditions.}  

We denote by  $\boldsymbol{n}_m$ and $\boldsymbol{t}_m$ the vectors normal and tangent to the membrane surface respectively (Fig.\ref{fig:sketch}).  We assume that the actin gel polymerizes at its natural density; $\rho =\rho^*$ at  the membrane, located at $z=u(x)$. Polymerization generates a normal actin velocity
$\boldsymbol{v}\cdot\boldsymbol{n}_m=(v_p+\delta v_{p}+\partial_t u /\sqrt{1+(\partial_x u)^2})$, where $\delta v_{p}$
accounts for small, protein regulated local modulations of actin polymerization \cite{Mihali2025}. 

The interfacial mechanics includes a frictional coupling between the gel and the membrane $(\boldsymbol{\sigma}\cdot\boldsymbol{n}_m)\cdot\boldsymbol{t}_m = \xi\boldsymbol{v} \cdot \boldsymbol{t}_m$, with $\xi$ the actin-membrane friction coefficient.  

In the incompressible limit, the gel velocity and therefore the actin flux reach zero at a particular distance from the surface, equal to $h_0=v_p/k_d$ for a flat surface, and modulated periodically for a sinusoidal surface \cite{Mihali2025}. As there are no sources of negative actin flux away from the membrane, the surface where the flux vanishes defines the free surface of the actin gel, beyond which the actin density must vanish. The location of the free gel surface on a corrugated membrane is fully determined by the conditions of vanishing normal flux and vanishing normal and tangential stresses, see \cite{Mihali2025}. 
The same approach is implemented here using the finite element method (FEM - code available on \href{https://github.com/dworthmuller/ActinWavyMembrane}{\underline{\blue{GitHub}}}) for finite compressibility, as described in Appendix \ref{appendix-fem}.

\subsection{Small deformation limit}

In order to analitically analyze the response of a viscous gel to membrane deformations, we perform a linear expansion of the velocity and density fields for small deformations ($|\nabla u|\ll1$) around their values $\rho_0(z)$ and $\boldsymbol{v}_0=v_{z0}\boldsymbol{e}_z$ corresponding to a flat interface, first neglecting all terms of order $\mathcal{O}(|\nabla u|^2)$: $ \boldsymbol{v} = \boldsymbol{v}_0 + \delta \boldsymbol{v}, \boldsymbol{\sigma} = \boldsymbol{\sigma}_0 + \delta \boldsymbol{\sigma} ,\rho = \rho_0 + \delta \rho$. As said above, all functions are expanded in Fourier modes assumed to vary along $x$, e.g., $\delta v_x=\delta v_{x,q} e^{iqx}$, etc. These modes are independent at the linear order.

The normal and tangent vectors read
\begin{align}
    \boldsymbol{n}_m &\approx (-\partial_x u_q\, \boldsymbol{e}_x + \boldsymbol{e}_z)\;, \\
    \boldsymbol{t}_m &\approx (\boldsymbol{e}_x + \partial_x u_q\, \boldsymbol{e}_z)\;.
\end{align}
The projection of the Stokes equation along $x$ and $z$ and the steady-state mass conservation read:
\begin{align}
    -iq\chi \delta \rho - 2 q^2\eta \delta v_x + \eta \partial_z^2 \delta v_x + iq \eta \partial_z \delta v_z &=0\;,\label{x-component1}\\
    -\chi \partial_z \delta \rho -  q^2\eta \delta v_z + 2\eta \partial_z^2 \delta v_z + iq \eta \partial_z \delta v_x &=0\;,\label{z-component1}\\
    iq\rho_0 \delta v_x + \rho_0\partial_z \delta v_z + \delta v_z \partial_z\rho_0 + \delta\rho\partial_z v_0\notag\\ + v_0 \partial_z\delta\rho = -kd\delta\rho\;.\label{mass-cons1}
\end{align}
Equations \ref{x-component1}, \ref{z-component1} and \ref{mass-cons1} define a linear problem for $(\boldsymbol{\delta v}, \delta \rho)$. Our goal is to determine the normal stress components evaluated on the membrane. 


\section{Steady state actin layer.}\label{section2}
The steady-state density and velocity profiles and the associated membrane stress can be computed analytically in the linear regime in the limit of infinite or vanishing compressibility: $\chi=\infty$ and $\chi=0$. These limits
provide explicit scaling laws for the normal membrane stress and serve as a benchmark for the the FEM simulations.
For finite compressibility the density and the flow are coupled even at linear order, preventing closed-form solutions. This motivates a numerical exploration of the biologically relevant  intermediate regime.

\subsection{Incompressible limit.} 
\begin{figure}[b]
    \centering
    \includegraphics[width=\linewidth]{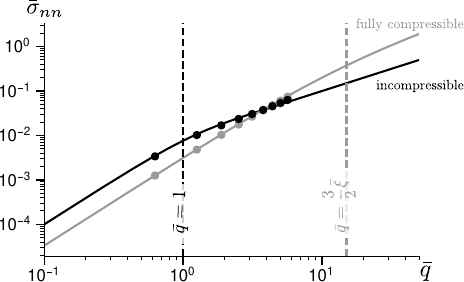}
    \caption{Log-log plot of the normal stress for $\delta v_{p,q}=0$ as a function of the dimensionless wavenumber $\bar q=qh_0$, in the high gel-membrane friction limit ($\bar\xi=10$). In the incompressible limit (Eq.~\ref{stress_incompressible}, black), 
    the stress exhibits two regimes controlled by the nondimensional parameter $q h_0$: a long wavelength scaling, $\sigma_{nn, q}/2\eta k_d \sim  {\bar q}^2$ for $\bar q \ll 1$ and a short wavelength scaling, $\sigma_{nn, q}/2\eta k_d \sim  {\bar q}$ for  ${\bar q} \gg 1$. The crossover is highlighted by the black vertical dashed line $\bar q =1$. In the fully compressible limit (Eq.~\ref{scaling_compressible_linear}, grey),    an additional friction controlled scale enters, $\bar\xi = \xi h_0 /2 \eta$. As a result the curve follows the same scalings as the incompressible case (${\sim\bar q}^2$ and $\sim\bar q$), but the crossover is now located at    $ \bar q = \frac{3}{2} \bar \xi$. Dots are the results of FEM numerics (Appendix \ref{appendix-fem}) in both cases. }
    \label{fig:scalings}
\end{figure}

The case of an incompressible viscous actin gel was derived in \cite{Mihali2025} and is merely recalled below.
The density is constant and equal to $\rho^*$, and the velocity field away from a flat surface is $\boldsymbol{v_0}=v_p(1-z/h_0)\boldsymbol{e}_z$ where $h_0=v_p/k_d$ is the gel thickness fixed by actin turnover. In that framework, the normal stress exerted by a polymerizing actin layer on a sinusoidal deformed membrane can be obtained analytically and does not depend on the friction parameter $\xi$ at linear order:
\begin{equation}
    \sigma_{nn, q} =-2\eta q (k_d u_q +\delta v_{p,q})\tanh\left(q h_0\right)\;.\label{stress_incompressible}
\end{equation}
In the absence of polymerization modulation ($\delta v_p=0$), the normal stress exerted by the gel acts to reduce the membrane deformations. It is negative (pushing) at the hills and positive (pulling) at the valleys of the deformed membrane, stabilising the flat membrane shape. It exhibits two different scaling laws depending on the ratio of gel thickness over deformation length scale $q h_0$: 
\begin{align}
    \lim_{qh_0 \rightarrow 0}\sigma_{nn,q} &= -2\eta k_d h_0 q^2 u_q\sim\eta v_p C_q\;,\nonumber\\
    \lim_{qh_0 \rightarrow \infty}\sigma_{nn, q} & = -2\eta k_d q u_q\;.
    \label{scaling_incompressible}
\end{align}
The stress is proportional to the local membrane curvature $C_q=-q^2 u_q$ in the large wavelength limit, and proportional to $|q|u_q$ in the small wavelength limit. 
These two scaling regimes arise because actin flow is modulated across the entire gel thickness in thin layers ($q h_0<1$), whereas in thick layers ( $q h_0>1$) it is confined to a boundary layer of depth $\sim 1/q$. Both limits follow intuitively from the role of incompressibility, which generates an isotropic negative pressure — effectively a suction — throughout the gel, equal to $P=-2\eta k_d$ for a flat membrane \cite{Mihali2025}. The normal stress $\sigma_{zz}$ vanishes because of the zero stress condition at the free surface, but there is a tangential stress $\sigma_{xx}=-P$. On a curved surface, the normal stress is essentially given by a Laplace law: $\sigma_{nn}=-PlC_q$, where $l\sim{\rm Min}[h_0,1/q]$ is the thickness over which the curvature is felt. This behaviour, shown in Fig.\ref{fig:scalings} (black curve), agrees with the numerical solution of the equations using FEM (black dots in Fig.\ref{fig:scalings}), described in Appendix \ref{appendix-fem}, in the limit of small deformation.

\subsection{Fully compressible limit.}\label{fully_compressible_limit} {The Stokes equation can be solved analytically for a fully compressible actin gel ($\chi =0$). The stress tensor Eq.\ref{stress} now reads $\boldsymbol{\sigma} =\eta (\nabla \boldsymbol{v} + \nabla\boldsymbol{v}^{\text{T}})$ and mass conservation is given by Eq.\ref{eq:mass-conservation-compressible}. The Stokes equation no longer depends on the density and we use the same boundary conditions as stated above.
At $0^{\text{th}}$ order (flat membrane) the force balance reduces to $\partial_z \sigma_{zz} = 2\eta\partial_z^2 v_z^0=0$ . Adding the condition of vanishing stress far from the boundary ($z\rightarrow\infty$) leads to a constant velocity $v_z^0 = v_p$ and a density that decays exponentially over the characteristic length $h_0=v_p/k_d$, according to $\rho_0=\rho^*e^{-z/h_0}$.

For the wavy membrane, we obtain the analytical expressions of the normal and tangential stress at $1^{st}$ order in $|\nabla u|$:
\begin{align}
   \sigma_{nn}(x,z)&=\frac{2\eta q}{4q\eta +3\xi} \, e^{iqx-qz}\Bigg(-v_p q u_q \xi(1-qz)\nonumber\\
   & + \delta v_{p,q}\Big(2\eta q (1+qz) + \xi (2+qz)\Big)
   \Bigg)
   \label{normal_stress_friction}\;,\\
    \sigma_{nt}(x,z)&= \frac{iq\eta}{4q\eta +3\xi}\, e^{iqx-qz} \Bigg(u_q v_p(qz-4)\xi\nonumber\;\\
    &+2\delta v_{p,q} (2q^2z\eta +\xi (1+qz))\Bigg)\;.
\end{align}

The normal stress calculated on the membrane surface reads
\begin{equation}
    \sigma_{nn}(x)\vert_{z=0} = -\frac{2q\eta(q u_q v_p \xi + 2 \delta v_{p,q} (q\eta +\xi))}{4 q\eta +3\xi}e^{iqx}\;.\label{stress_compressible}
\end{equation}
From this expression we obtain the scaling for large and small wavelength limits for $\delta v_p =0$:
\begin{align}
    \lim_{\eta q/\xi \rightarrow 0}\sigma_{nn,q} &= -\frac{2}{3} \eta v_p q^2 u_q \sim \eta v_p C_q\;,\nonumber\\
    \lim_{\eta q/\xi \rightarrow \infty}\sigma_{nn, q} &= -\frac{1}{2}\xi v_p q u_q \;.
    \label{scaling_compressible_linear}
\end{align}

It is interesting to compare the normal stress for incompressible and fully compressible gels (Eqs.~\ref{scaling_incompressible} and \ref{scaling_compressible_linear}). Both exhibit the same scalings in the small ($\sigma_{nn,q}\sim q^2 u_q$) and large ($\sigma_{nn,q}\sim q u_q$) wave length limits, but the crossover wavelengths are distinct. It is set by actin turnover  for incompressible gels ($q h_0\sim1$), and by the ratio of viscosity to friction  in the fully compressible case ($q\eta/\xi\sim1$). Furthermore, while the expressions only differ by a numerical factor for small $q$, the stress from a fully compressible gel varies linearly with the gel-membrane friction parameter $\xi$ for large $q$, and therefore vanishes in the case of perfect slip, while the stress from an incompressible gel is independent of friction for any value of $q$.
Qualitatively, this difference stems from the fact that in a compressible gel, if tangential flows are not hindered by any friction, density can adjust freely to prevent the existence of shear stress that could arise from the gel flowing normally away from a corrugated membrane.
Finite friction resisting tangent slip is the only source of  stress transmission at the membrane. It does so through the boundary force balance condition
$(\boldsymbol{\sigma}\cdot\boldsymbol{n}_m)\cdot\boldsymbol{t}_m = \xi\boldsymbol{v}\cdot \boldsymbol{t}_m$, 
leading to Eq.\ref{normal_stress_friction}.

The fully compressible case can be used to benchmark the FEM solutions against the analytical solution in the limit $\chi \rightarrow 0$. The agreement is excellent, as shown in Fig.\ref{fig:analytics_vs_fem} for the stress component $\sigma_{zz}(z)$ as a function of the distance from the membrane. This quantity qualitatively shows the same behavior as for the incompressible fluid (at high friction) with a change of sign at $q z=1$ (Eq.\ref{normal_stress_friction}) and an exponential decay over $1/q$ if $qh_0\gg1$.

\begin{figure}[t]
    \centering
    \includegraphics[width=\linewidth]{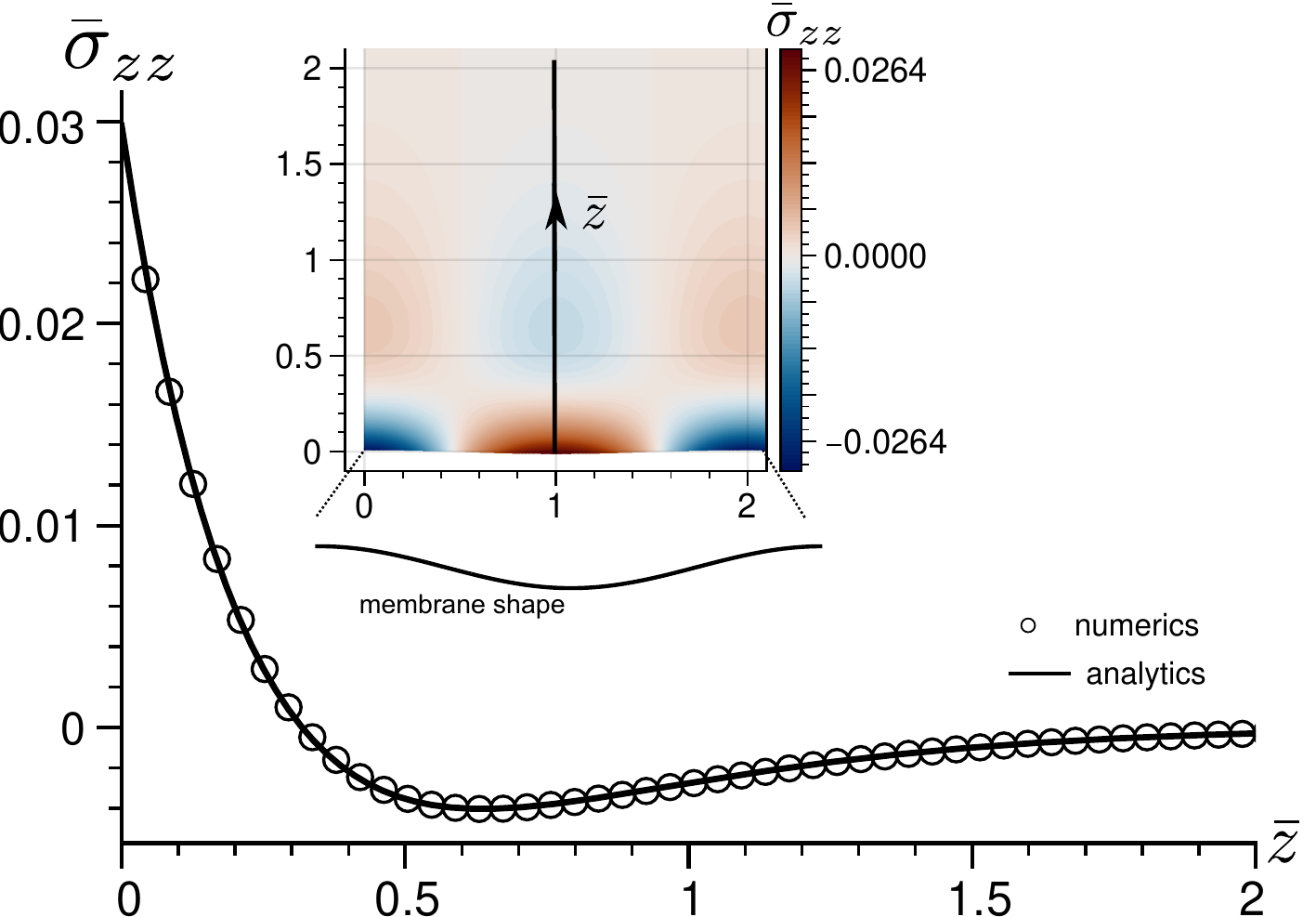}
    \caption{Comparison of FEM and analytical predicted stress tensor component $\sigma_{zz}(z)/2\eta k_d$ (corresponding to $\sigma_{nn}$, Eq.\ref{normal_stress_friction}, at the linear level) along the slice (black line) located at a minimum on the membrane deformation ($qx=\pi$). We find near perfect agreement between the simulations and analytics. 
    The parameters are chosen as $\  \chi/2\eta k_d =0,\ \rho^* =1,  \xi h_0/2\eta =20,  q h_0 =\pi$. Numerical stability is ensured by adding a small diffusive contribution to the transport equations (see Appendix \ref{appendix-fem}) with
    $D/k_dh_0^2=0.01$. All parameters are non-dimensional.}
    \label{fig:analytics_vs_fem}
\end{figure}

\subsection{Finite compressibility.} Obtaining a fully analytical solution of the first order equations Eqs.\ref{x-component1},\ref{z-component1},\ref{mass-cons1} is challenging, because the linear perturbations couple the flow and the density variations. The pressure-like contribution $-\chi(\rho - \rho^*)$ is no longer uniform and the viscous stresses inherit an explicit $x$-dependency through both $\boldsymbol{v}(x,z)$ and $\rho(x, z)$. This dependence must be $e^{i q x}$ at the linear level.  Consequently, even at $1^{\text{st}}$ order in membrane deformation ($q u_q\ll 1$), closed-form expressions are not available. We therefore rely on FEM to obtain the stress, velocity and density fields and to explore the full $(\xi, \chi)$ parameter space. The full non-dimensionalization, weak formulation and boundary condition implementation are provided in Appendix \ref{appendix-fem}.
By contrast, the $0^{\text{th}}$ order problem is tractable analytically. At this order, Eq.\ref{z-component1} and Eq.\ref{mass-cons1}, are invariant along $x$ and Stokes equations reduce to a one-dimensional problem. The membrane normal and tangent vectors reduce to $\boldsymbol{n}_m=\boldsymbol{e}_z$ and $\boldsymbol{t}_m=\boldsymbol{e}_x$ and the stress slip condition simplifies to a vanishing stress $\Big((\boldsymbol{\sigma}\cdot\boldsymbol{n}_m)\cdot\boldsymbol{t}_m\Big)\vert_{z=0}=\xi \boldsymbol{v}_t\vert _{z=0}=0$.
At the membrane, density satisfies $\rho_0\vert_{z=0}=\rho^*$ while in the mass conservation law, the turnover $k_d$ is balanced by the advective flux set by the imposed normal polymerization velocity, $\boldsymbol{v}\vert_{z=0}\cdot \boldsymbol{n}_m=v_p$ (discussed in Appendix \ref{0-solutions} and shown in \cref{fig:flat}).
The solution for the reference density profile and velocity follows the same baseline procedure used in the "active wetting layer" approach of \cite{Joanny2013ActiveWetting}. Here, we do not analyze the wetting transition, but rather use the flat state solution as a reference state to study the shape instabilities of the membrane: we perturb around this state, solve the resulting Stokes problem and compute the induced stresses on the membrane that control the stability of the membrane corrugations. 

Fig.\ref{fig:stressmap} shows the results of the FEM in the friction - compressibility ($\xi,\chi$) parameter space for a given value of $q$. The results smoothly interpolate between the fully compressible (Eq.\ref{stress_compressible}) and incompressible (Eq.\ref{stress_incompressible}) solutions. The dimensionless compressibility $\bar\chi=\chi/(2\eta k_d)$ can be defined from the equations. This can be understood as the inverse of the product of a viscoelastic time scale obtained as the ratio of shear viscosity over bulk compressibility, and the actin turnover rate. As discussed above, the dimensionless friction $\bar\xi = \xi h_0/2\eta$ compares the gel thickness with the typical friction length scale $\eta/\xi$. The relevance of the latter can be gathered from the analytical expression in the fully compressible case Eq.\ref{scaling_compressible_linear}, which shows that frictional effects are relevant for $\bar\xi\gtrsim\bar q=qh_0$. 
Fig.\ref{fig:stressmap} shows that the incompressible regime is approached for a dimensionless compressibility $\bar\chi=\chi/(2\eta k_d) \simeq 40$. for $qh_0\simeq1$.  Appendix \cref{app:compressible} and \cref{fig:scalings_nonlinear}  show that the effect of a finite compressibility are more relevant for larger $q$, especially in the case of small friction, where non-linear effects become important.

\begin{figure}[t]
     \centering
     \includegraphics[width=\linewidth]{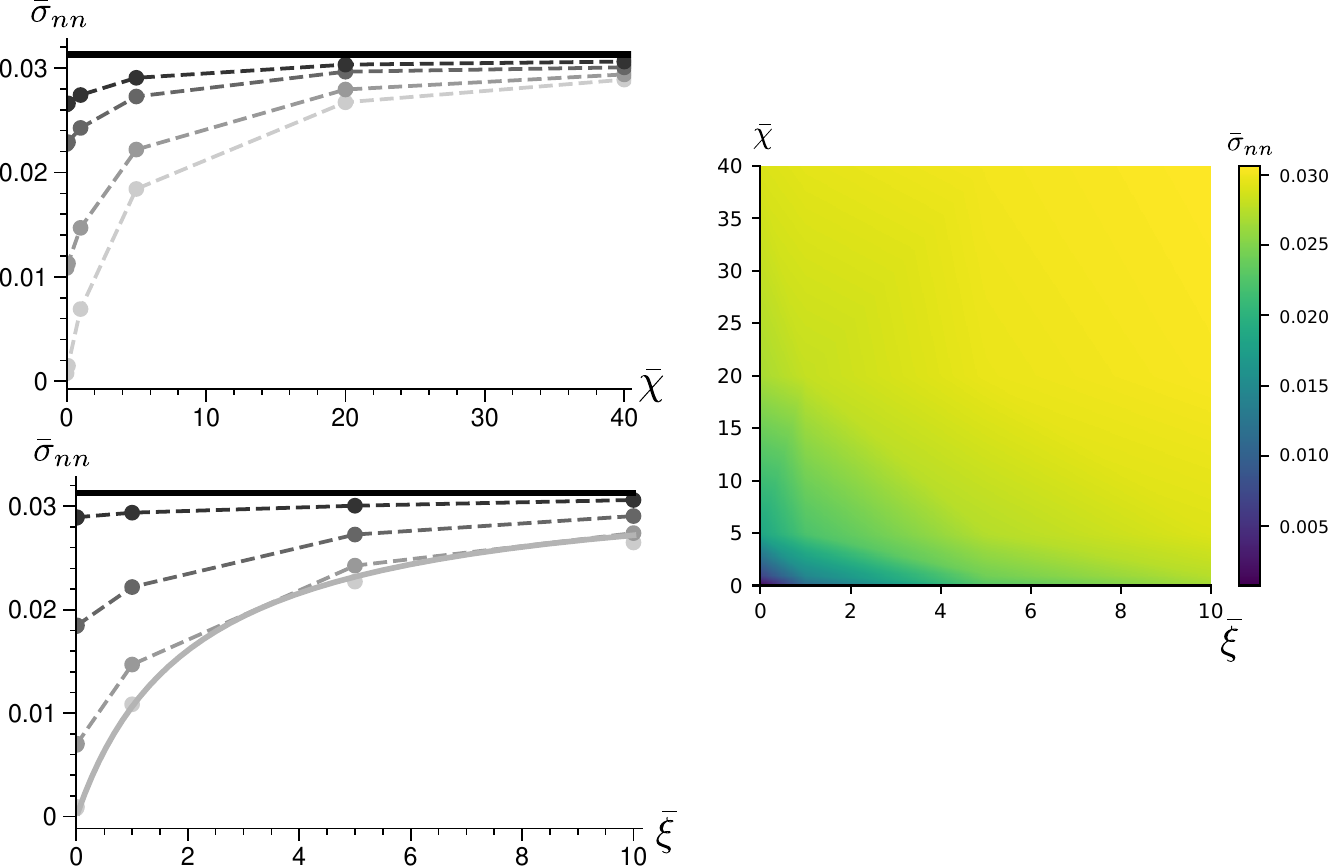}
     \caption{Stress distribution on the membrane as a function of friction $\xi$ and compressibility $\chi$. The color scale indicates the magnitude of the stress variation over the phase space. The dimensionless compressibility is $\bar\chi=\chi/(2\eta k_d)$ and the dimensionless friction is $\bar\xi = \xi h_0/2\eta$.
     The two side plots show the stress as a function of the friction for different values of the compressibility, $\bar\chi=[0,1,5,40]$, and the stress as a function of the compressibility for different values of the friction parameters, $\bar \xi =[0,1,5,10]$. Bold curves show the incompressible (black, \cref{stress_incompressible}) and fully compressible (grey, \cref{stress_compressible}) limits at the linear level. Dashed lines are provided purely to guide the eye between data points. Non-linear effects are not apparent on this plot. They are discussed in the appendix and shown in \cref{fig:incompressible}. 
     Parameters: $\bar u_0=0.01, h_0q=\pi.$}
     \label{fig:stressmap}
 \end{figure}

\section{Coupling actin and membrane dynamics.}
A hallmark of active cell mechanics is the existence of feedback between local shape and active stress. At the cell membrane, these feedbacks are often mediated by the recruitment of curvature-sensitive proteins that affect actin dynamics \cite{Zhao:2011, lou2019, tsai2022}. In \cite{Mihali2025}, we showed that proteins such as I-Bar domain proteins that enhance actin polymerization in membrane regions curved away from an incompressible gel ($\chi\rightarrow\infty$) may render the flat membrane linearly unstable if their recruitment is fast enough. Here we extend this analysis to finite compressibility and we account for the effect of actin-membrane friction. 

Spatial variations of actin assembly at the membrane can generate stresses, even in the absence of membrane corrugations. This stress is finite even in the case of perfect slip ($\xi=0$) and reads $\sigma_{nn,q}=-\eta q\delta v_{p,q}$ for a fully compressible gel (Eq.\ref{stress_compressible}) and $\sigma_{nn,q}=-2\eta q\delta v_{p,q}\tanh{qh_0}$ (Eq.\ref{stress_incompressible}) for  an incompressible gel.
If the modulation of actin polymerization is controlled by curvature-sensitive proteins that are fast to adjust to the local membrane shape, this is captured at the linear level by a simple relationship between polymerization velocity and curvature: $\delta v_p(x) = \alpha C=-\alpha q^2 u_q$, where $\alpha$ (of unit of m$^2$/s) is the coupling strength \cite{Mihali2025}.
Unlike the stress induced by membrane corrugation at uniform $v_p$, first term in Eq.\ref{stress_compressible}, the stress generated by the velocity modulations remains finite as $\xi \rightarrow 0$. This is because the stress is being transmitted at the membrane due to the nonuniform boundary actin flow. 

The resulting normal stress (denoted $\delta\sigma_{nn}$) takes the following forms in the fully compressible and incompressible limits
\begin{align}
  \chi\rightarrow0:\qquad \delta\sigma_{nn} &= \frac{4(q\eta+\xi)}{4q\eta+3\xi}\eta\alpha q^3 u_qe^{iqx}\;,\\
    \chi\rightarrow\infty:\qquad \delta\sigma_{nn} &=2\eta\alpha q^3\tanh{qh_0} u_qe^{iqx}\;.
\end{align}
Remarkably  in the fully compressible limit, it is independent of the actin turnover rate $k_d$ and shows little dependence on friction (at most a numerical factor $3/4$) and remains finite even for $\xi \rightarrow 0$.

To determine whether this mechanism mentioned above can spontaneously deform the membrane, we couple the actin-generated stresses with the membrane restoring forces through a local force balance $\sum \sigma_i = \sigma_{nn} + \sigma_m=0$. We use Monge parametrization to describe the shape of the membrane, $u(x)$ and write the Helfrich free energy of the membrane as \cite{helfrich1973}}
\begin{equation}
    F_{\text{Helfrich}} = \int \Bigg(\frac{\kappa}{2}(\nabla^2 u)^2 + \frac{\gamma}{2}(\nabla u)^2\Bigg) d^2x\;,
\end{equation}
where $\kappa$ is the membrane bending rigidity and $\gamma$ is the membrane tension. The membrane restoring stress is obtained from the functional derivative, $\sigma_m = \delta F_{\text{Helfrich}}/\delta u=-\kappa \nabla^4 u + \gamma \nabla^2 u$.  In Fourier space the membrane contributions becomes $\sigma_{m,q} =-(\kappa q^4+\gamma q^2)u_q$, and the total normal stress of the membrane for a compressible gel reads
\begin{equation}
   \sigma_{\text{tot, q}}= -2 q^2 u_q \eta\frac{v_p\xi - 2 q \alpha (q \eta + \xi)}{4 q \eta + 3 \xi}-(\kappa q^4+\gamma q^2)u_q
   \label{sigmatot}
\end{equation}
\begin{figure}[h]
    \centering
    \includegraphics[width=\linewidth]{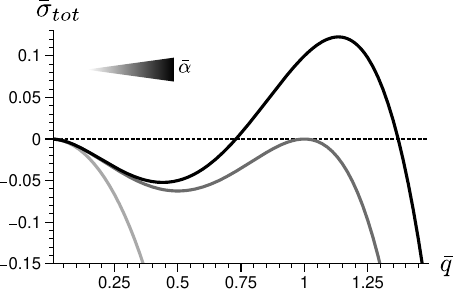}
    \caption{Total stress (gel contribution and membrane restoring forces) $\bar \sigma_{tot} = \sigma_{tot}\lambda ^2/\gamma u_q$ for a fully compressible gel ($\bar \chi=0$, \cref{sigmatot}) in the absence of actin-membrane friction ($\bar \xi=0$). The stress is shown as a function of $\bar q= q \lambda$ where $\lambda=\sqrt{\kappa/\gamma}$ is the membrane characteristic length, for increasing values of the curvature-polymerization coupling strength $\bar\alpha=\alpha \eta/(\gamma \lambda) = [0\ \rm{(light\ grey)}, 2\ \rm{(grey)}, 2.1\ \rm{(black)}]$. The membrane is linearly unstable when the total stress is positive.}
    \label{fig:instabilities}
\end{figure}

Extension to finite compressibility is conceptually straightforward, but requires the use of the numerical results discussed above and displayed in \cref{fig:stressmap}.

If $\alpha>0$, meaning that polymerization is promoted in membrane regions curved away from the gel, curvature modulation of actin polymerization introduce a positive contribution to the membrane stress that can render the flat membrane linearly unstable. 
In the frictionless limit, the force balance equation can be written as $\frac{\gamma u_q}{\lambda^2}(\bar{\alpha} (\lambda q)^3 - (\lambda q)^4 - (\lambda q)^2)=0$,
where $\lambda = \sqrt{\kappa/\gamma}$ is the membrane's characteristic length scale and $\bar\alpha=\eta\alpha/(\gamma\lambda)$. One can easily show that the total stress has the sign of the membrane deformation, destabilizing the flat shape, for a range of wavenumbers centered around $\lambda q=1$ if $\bar\alpha>2$. \cref{fig:instabilities} illustrates the transition from a stable to an unstable membrane as the coupling between polymerization and curvature, controlled by $\alpha$, is increased.

\section{Nonlinear contribution of the stress}\label{non_linear_derivations}
\begin{figure}[t]
    \centering
    \includegraphics[width=\linewidth]{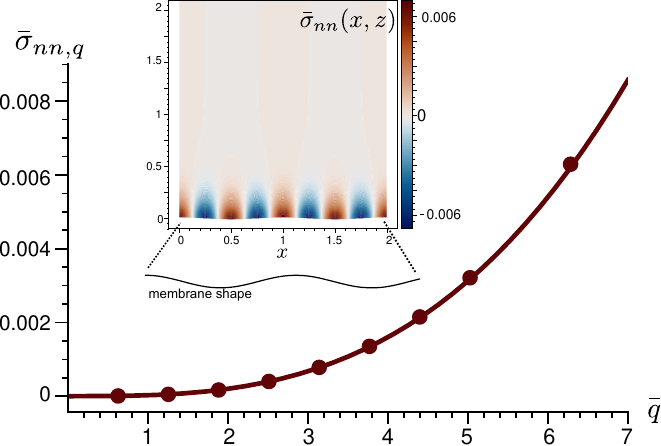}
    \caption{$2^{\text{nd}}$ order stress for a fully compressible frictionless gel ($\chi=0,\ \xi=0$). The main plot shows the normal stress amplitude $ \sigma_{nn,q}/2\eta k_d$ as a function of $q$ at $qx=0\ ({\rm mod}\ \pi)$; 
    solid line shows the analytical results and the data points denote the FEM results. 
   The inset shows the spatial stress distribution $\bar \sigma_{zz}(x,z)$
    for $q h_0=2 \pi$, capturing the period doubling associated with the nonlinear effects. Inset: $u_q/h_0=0.01$}
    \label{fig_nonlinear}
\end{figure}

In the absence of friction ($\xi=0$) the membrane stress generated by a fully compressible gel vanishes at $1^{\text{st}}$ order in deformation and the stress is dominated by higher order effects even in the limit of small deformation. While an arbitrary membrane shape cannot be decomposed into independent Fourier modes in this case, it is still possible and informative to compute $2^{\text{nd}}$ order effects for a sinusoidal membrane shape. If $\chi=0$, density variations decouple from the Stokes equation. The perturbative approach described above may be expanded to $2^{\text{nd}}$ order in the small quantity $qu_q$.
The $1^{\text{st}}$ order contribution, given by Eq.\ref{stress_compressible} follows the periodicity of the membrane shape $u_q e^{i q x}$. The $2^{\text{nd}}$ order contribution generates higher harmonics $\propto u_q^2 e^{2iqx}$.
In the absence of polymerization modulation  ($\delta v_p=0$), the gel stress on the membrane reads up to $2^{\text{nd}}$ order
\begin{align}
    \sigma_{nn} &=-\frac{2 q^2  v_p \eta \xi}{4q\eta + 3 \xi}u_q e^{iqx}\notag\\
    &+\frac{2 q^3  v_p \eta (8 q^2\eta ^2 - 4q\eta\xi  -3 \xi^2)}{(4q\eta + 3 \xi)(8q\eta + 3 \xi)}u_q^2 e^{2iqx}
    \label{eq:non-linear}\;.
\end{align}
The most direct test of the $2^{\text{nd}}$ order expansion is the frictionless limit when the $1^{\text{st}}$ order vanishes and the stress reduces to $\sigma_{nn} = \frac{1}{2}(q^3 v_p \eta) u_q^2e^{2iqx}$. The expression shows an excellent match with the numerical results for small deformation in Fig.\ref{fig_nonlinear}, both regarding the stress amplitude as a function of $q$ and the presence of second harmonics in the stress distribution.

For a fully compressible gel, the $2^{\text{nd}}$ order normal stress is positive (pulling) both at the hills and at the valleys of the deformation, destabilising the former and stabilising the latter. While a detailed analysis of non-linear effects is challenging and will be the subject of a subsequent publication, one may already envision that non-linear shape evolution of a deformable membrane will involve amplification of the deformed regions.

\section{Numerical values of parameters}
To estimate the relevance of the different regimes discussed in this paper in the context of a biological cell, the important dimensionless parameters are the non dimensional compressibility $\bar \chi = \chi/(2 \eta k_d)$ and friction $\bar \xi = \xi h_0/2\eta$. The relevance of friction also depends on the wavevector $\bar q=q h_0$.
Typical values of the different parameters from the literature are reported in \cref{tab:params}. The compressibility can vary widely depending on the experimental conditions, so that $\bar\chi$ can potentially cover a wide range of value $\sim 10^{-4}-10^2$, thereby spanning the entire range from fully compressible to incompressible gels. On the other hand, the dimensionless friction $\bar\xi$ is expected to be rather small. The friction coefficient can be estimated as the product of the membrane (2D) viscosity times the density of actin-membrane linkers: $\xi\simeq\eta_{2D}^{\rm memb}\rho_b$, with $\rho_b\simeq 1/(10-100 \mathrm{nm})^2$, yielding $\bar\xi\sim 10^{-5}-10^{-2}$. This suggests that a full-slip approximation is reasonable. In particular, the characteristic length scale defined by the ratio of gel viscosity over gel-membrane friction $\eta/\xi\sim 10-10^3\mathrm{\mu m}$ is of the order of the cell size or larger. This means that the scale of the stabilising stress in the compressible limit (\cref{sigmatot}), which is set by the friction parameter, is rather small ($\sim\xi v_p\lesssim 10$Pa).

\begin{table}[t]
\centering
\renewcommand{\arraystretch}{1.2}
\begin{tabular*}{0.8\columnwidth}{@{\extracolsep{\fill}}lll}
\toprule
\textbf{Parameter} & \textbf{Value} & \textbf{Ref.} \\
\midrule
$k_d$ & $0.1\,\mathrm{s^{-1}}$ & \cite{fritzsche2013, kuhn2005} \\
$v_p$ & $0.01\,\mu\mathrm{m/s}$ & \cite{pollard2003,bieling2016} \\
$h_0$ & $0.1\,\mu\mathrm{m}$ (derived) & -- \\
$\eta$ & $10^{4}\,\mathrm{Pa\,s}$ & \cite{salbreux2012,charras2008} \\
$\chi$ & $1$--$10^{6}\,\mathrm{Pa}$ & \cite{Xu1998,Cordes2020,FischerFriedrich2016} \\
$\eta_{2D}^{\mathrm{memb}}$ & 
$10^{-9}-10^{-6} \mathrm{Pa.m.s}$ & \cite{salbreux2012,Zgorski2019,Fitzgerald:2023}\\
$\xi$ & 
$10^6-10^9\,\mathrm{Pa.s/m}$ & \cite{Zgorski2019}, see text \\
$\eta/\xi$ & $10-10^3\mathrm{\mu m}$ & derived \\
\bottomrule
\end{tabular*}
\caption{Expected values for the physical parameters.}
\label{tab:params}
\end{table}

\section{Discussion}
In this paper, we provide analytical expressions for the stress exerted by an actin gel polymerizing on a corrugated membrane. We extend earlier work \cite{Mihali2025}, which demonstrated that an incompressible gel may render a flat membrane linearly unstable to spontaneous deformations,  provided appropriate feedback between local membrane curvature and actin polymerization. Here we show how the mechanical properties of the actin gel, and in particular its compressibility and dynamical coupling  with the membrane (friction), modify this picture. \\
The mechanical stress exerted by the active gel on a corrugated (sinusoidal) membrane can be decomposed into two contributions,  additive at the linear level in membrane deformation. The first comes from the uniform growth of a gel on a non-flat surface, and the second from the non-uniform growth of a gel modulated by the local membrane curvature. The latter is responsible for the instability, and shows little influence from the gel compressibility and gel-membrane friction in the range of friction expected in cellular systems and if $qh_0\gtrsim 1$. The former, which stabilises the flat membrane shape at the linear level, is very much affected by the gel compressibility. For an incompressible gel, depolymerization in the bulk of the gel creates a suction effect which gives rise to a tension in the direction parallel to the membrane. This tension acts to suppress membrane undulations via a Laplace pressure-like effect which does not rely on gel-membrane friction. As a consequence, the stabilising and destabilising stress terms scale similarly with the gel mechanical properties (its viscosity $\eta$, \cref{stress_incompressible}) and the existence of a range of unstable wave lengths requires a rather strong coupling between gel polymerization and local membrane curvature: $\delta v_p(q)>v_p$ \cite{Mihali2025}. For a fully compressible gel, the suction effect is absent, and the scale of the stabilising normal stress is controlled by gel-membrane friction (\cref{stress_compressible}).
Compared to the incompressible case, the stabilising term is potentially much smaller and vanishes entirely in the absence of friction. The instability is therefore only opposed by the mechanics of the membrane (\cref{sigmatot}), and can be expected to exist over a broader range of wave length. Numerical estimates of typical friction parameter values in the cellular context support the concept of a curvature instability and spontaneous cellular protrusions, driven by curvature-dependent modulation of the polymerisation dynamics of a compressible actin gel at the cell membrane.

\begin{acknowledgments}
This project has received funding from the European Research Council (ERC) ERC-SyG (Grant agreement ID: 101071793)
\end{acknowledgments}

\appendix
\section{Contribution of bulk viscosity}\label{2-viscosity}
\appendixfigures
In the main text, we set bulk viscosity $\eta_b=0$ and model the compressible actin gel with the constitutive maw $\boldsymbol{\sigma}=-P(\rho)\boldsymbol{I}+\eta(\nabla v + \nabla v^{T})$. For completeness, we now consider the general viscous constitutive relation as 
\begin{equation}
    \boldsymbol{\sigma}=(-P(\rho + \eta_b \nabla\cdot \boldsymbol{v})\boldsymbol{I}+\eta(\nabla v + \nabla v^{T})\;.
\end{equation}
We solve the Stokes and mass conservation equations with the modified stress tensor for the fully compressible limit in the same boundary conditions as in Section \ref{BC}. The presence of $\eta_b$ affects the isotropic part of the viscous stress and enters the solutions with a modified prefactor
\begin{equation}
    \sigma_{nn} = -\frac{2q^2 u_q v_p \eta^2 \xi}{2q\eta(2\eta +\eta_b)+(3\eta +\eta_b)\xi};\,
\end{equation}
the scaling with $q$ remains unchanged to Eq. \ref{stress_compressible} of the main text. Introducing the bulk viscosity changes only the prefactor of the stress, $\sim(\eta + \eta_b)$, but does not change the overall behaviour of stress and the way it depends in wavelength and also friction. 

\section{Gel with finite compressibility growing on a flat surface}\label{0-solutions}
\begin{figure}[b]
    \centering
    \includegraphics[width=\linewidth]{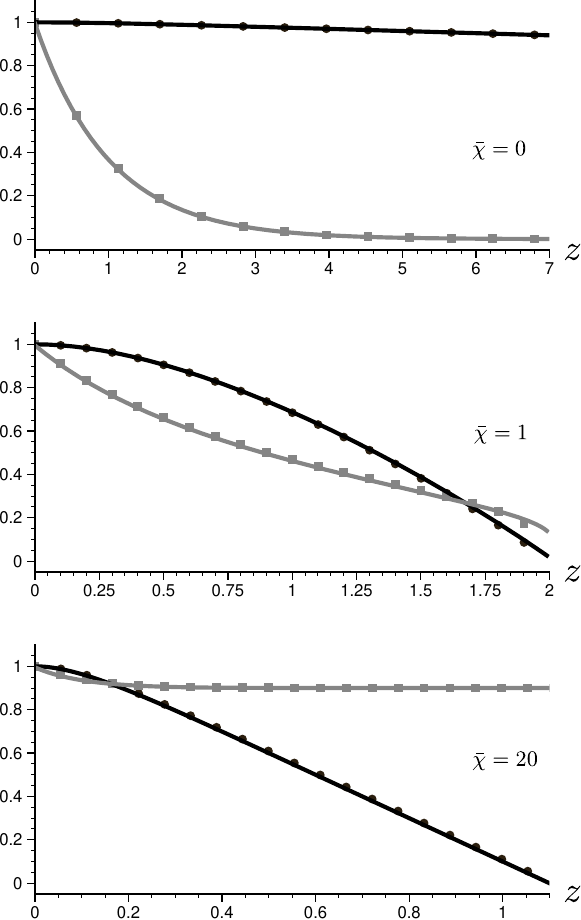}
    \caption{Compressible flow near a flat membrane:Velocity field $v_0(z)$ (grey) and density field $\rho_0(z)$ (black)  as a function of the distance to the membrane $z$ for different values of $\bar\chi=\chi/(2\eta k_d)$ (with $\rho^*=1$). Lines represent analytical solutions and dots/squares the corresponding finite element solutions.}
    \label{fig:flat}
\end{figure}

\appendixfigures
The density and velocity profiles of a gel with finite compressibility growing on a flat surface constitute the zeroth order solution of our expansion and can be obtained analytically in a way similar to the one proposed in \cite{Joanny2013ActiveWetting}. 
All the fields depend only on the distance z from the membrane. 
The velocity field is purely normal, $\boldsymbol{v}_0 = v_0(z) \boldsymbol{e}_z$ and the density profile $\rho_0(z)$ satisfies $\partial_z (\boldsymbol{v}_0 \rho_0)=-k_d\rho_0$ and $-\partial_z P_0 + 2\eta\partial_z^2 \boldsymbol{v}_0=0$, where $P_0=\chi(\rho_0-\rho^*)$. At the membrane, $z=0$, we impose the polymerization to be $v_0(0)=v_p$ and density $\rho_0=\rho^*$
The analytical expressions are cumbersome and not given explicitly here. The zeroth order fields $v_0(z)$ and $\rho_0(z)$ are shown on Fig.\ref{fig:flat}, which compares the analytical solutions of the velocity and density profiles with finite element simulations, for several values of compressibility going from the fully compressible limit to the incompressible limit. The agreement is excellent over the full range of parameter $\chi$.

\section{Finite Element Method for Compressible Flow}\label{appendix-fem}
\appendixfigures
To explore the full ($\xi, \chi$) parameter space, we solve the Stokes equation numerically using a finite element method. To derive the non-dimensional form of the Stokes equation and the mass-conservation law, we introduce characteristic scales that reflect the physical geometry and dynamical timescales of the system. The characteristic stress/pressure scale $p_s =2\eta k_d$, where $\eta$ is the dynamic viscosity, the velocity scale $v_s =v_p$ with length scale $h_0$, and the natural timescale which follows is $t_s=1/k_d$.
In simulations the wavelength of membrane corrugations is set by $q = m\;2\pi/L_x$, with $m \in \{0,1,2,3,...\}$ and $L_x$ the simulation domain size in $x$-direction. Using this definitions, the governing equations yield the non-dimensional form
\begin{align}
   -\bar \chi \bar \nabla \bar \rho + \frac{1}{2}\bar{\nabla}^2 \boldsymbol{\bar v}+\frac{1}{2}\bar{\nabla}(\bar{\nabla}\cdot \boldsymbol{\bar v}) & = 0 \;,\label{stok_eq_nondim}\\
   \bar \nabla \cdot(\bar \rho \boldsymbol{\bar v}) - \bar D \bar{\nabla}^2 \bar \rho + \bar \rho & = 0 \label{mass_cons_nodim}\;.
\end{align}
where an additional diffusive contribution is included in the mass-conservation law to enhance numerical stability of the finite element discretization (discussed later in the text). The dimensionless control parameters are $\bar D= D/(k_d h_0^2)$ and $ \bar \chi = \chi  /(2 \eta k_d)$.

The boundary conditions simplify accordingly. On the membrane surface $\partial\Omega_m$ the normal velocity is prescribed as
\begin{align}
    \bar v_n &= \boldsymbol{\bar v \cdot n}\vert_{z=\partial \Omega_m}=1
\end{align}
while the tangential component satisfies a Robin-type condition
\begin{align}
    \bar v_t &= \boldsymbol{\bar v \cdot t}\vert_{z=\partial \Omega_m}=\bar \sigma_{nt}/\bar \xi \;, \label{RobinBC}
\end{align}
with $\bar \xi = \xi h_0/2\eta$ representing a dimensionless interfacial friction coefficient where $h_0$ is the actin cortex thickness. 

\subsection{Weak formulation}

To derive the weak formulation, the Stokes Eq.\ref{stok_eq_nondim} is multiplied by a vector-valued test function $\boldsymbol{w} \in \mathcal{V}_{V}$  and the mass-conservation equation is multiplied by a scalar-valued test function $w_s \in \mathcal{V}_{S}$, where $\mathcal{V}_{V}$ and $\mathcal{V}_{S}$ denote the spaces of vector- and scalar-valued test functions on $\Omega$. After integration by parts, the momentum balance yields 
\begin{align}
    F_{\boldsymbol{v}} &= \int_{\Omega} (\bar \nabla \cdot \boldsymbol{\bar \sigma}) \cdot \boldsymbol{w} d\Omega\\ &= \int_{\partial \Omega} (\boldsymbol{\bar \sigma}\cdot \boldsymbol{N})\cdot \boldsymbol{w} \; ds
    - \int_{\Omega} \boldsymbol{\bar \sigma} \colon \bar \nabla \boldsymbol{w} \; ds\\ & =0 \;,
\end{align}
 where $\boldsymbol{N} = -\boldsymbol{n}$ denotes the outward pointing normal on the boundary and $:$ denotes double tensor contraction. Traction boundary conditions are imposed naturally through the boundary integral. On boundaries where zero traction is prescribed, this contribution vanishes. It also vanishes on Dirichlet segments, 
where the test function is chosen to vanish. The boundary integrals over the periodic boundary $\partial \Omega_{\mathrm{pb}}$ cancel in pairs since the solution and test functions coincide on paired faces while the outward normals have opposite orientation. We impose zero stress at the free surface $\partial \Omega _{f}$. As a consequence, the only non-zero traction contribution arises on $\partial \Omega_m$. Splitting the traction into normal and tangential  components gives
\begin{align}
    \int_{\partial \Omega_m}(\boldsymbol{\bar \sigma}\cdot \boldsymbol{N})\cdot \boldsymbol{w} \; ds &= -\int_{\partial \Omega_m}(\boldsymbol{\bar \sigma}\cdot \boldsymbol{n})\cdot \boldsymbol{w} \; ds \notag\\
    &=-\int_{\partial \Omega_m}\bar\sigma_{nn}w_n \;ds - \int_{\partial\Omega_m}\bar \sigma_{nt}w_t \;ds \;,
\end{align}
with $w_n = \boldsymbol{w\cdot n}$ and $w_t = \boldsymbol{w \cdot t}$. The tangential traction is replaced by $\bar \sigma_{nt} = \bar \xi \bar v_t$, consistent with the Robin condition and the sign convention associated with the inward normal i.e. tangential flow causes friction in the direction of the flow. 
The normal velocity $\bar{v}_n = 1$ is enforced weakly by means of a pure penalty formulation. 
To this end, we add the boundary functional
\begin{equation}
    \frac{\beta}{2\Delta}
\int_{\partial\Omega_m} (\bar{v}_n - 1)^2\, \diff s ,
\end{equation}
whose first variation yields the additional contribution
\begin{equation}
F_{v,\mathrm{pen}} =
\frac{\beta}{\Delta}
\int_{\partial\Omega_m} (\bar{v}_n - 1) w_n\, \diff s .
\end{equation}
Here $\Delta$ denotes the average mesh size and $\beta>0$ controls the strength of the enforcement. Combining all contributions yields the weak form of the momentum balance
\begin{align}
   & F_v = -\int_{\Omega}\boldsymbol{\bar \sigma \colon}\bar \nabla \boldsymbol{w} \; ds - \int_{\partial\Omega_m}\bar\sigma w_n \;ds - \int_{\partial\Omega_m}\bar \xi \bar{v}_t w_t\; ds \notag\\
    &  +\frac{\beta}{\Delta}\int_{\partial\Omega_m}(\bar{v}_n - 1)w_n \;ds=0\;.
\end{align}
The weak formulation for the mass-conservation follows from applying the divergence theorem and integrating by parts. After rearrangement, one obtains
\begin{align}
    F_{\rho} &=  \int_{\Omega}\bar \nabla \cdot (\bar \rho \boldsymbol{\bar v}) w_s \; d\Omega+ \int_{\Omega} \bar \rho w_s\;d\Omega - \int_{\Omega}\bar D (\bar{\nabla}^2 \bar \rho)w_s \; d\Omega \notag \\ 
    &= \int_{\Omega} \bar \rho w_s\;d\Omega +\int_{\partial\Omega} (\bar \rho \boldsymbol{\bar v} - \bar D \bar \nabla \rho ) \cdot \boldsymbol{N}w_s \;ds \notag \\ &- \int_{\Omega}(\bar \rho \boldsymbol{\bar v}) \cdot \bar \nabla w_s \;d\Omega +\bar D\int_{\Omega}\bar \nabla \bar \rho \cdot \bar \nabla w_s \;d\Omega\;.
\end{align}
The boundary integrals over $\partial \Omega_{pb}$ vanish due to the same argument as mentioned above. We keep the term for the outflux boundary (which means that we do not specify the outflux here). 
On the membrane boundary $\partial \Omega_m$, we enforce
\begin{equation}
    \bar \rho \vert_{\partial \Omega_m}  =\bar \rho^{*}\;,
\end{equation}
through a Dirichlet boundary condition. Consequently, the test function $w_s$ vanishes. Collecting all contributions leads to the final weak formulation,
\begin{align}
    &F_{\rho} =\int_{\Omega}\; \bar{\rho} \;w_s\;d\Omega \notag\\ & +\int_{\partial\Omega_f} (\bar \rho \boldsymbol{\bar v} - \bar D \bar \nabla \rho ) \cdot \boldsymbol{N}w_s \;ds - \int_{\Omega}(\bar \rho \boldsymbol{\bar v}) \cdot \bar \nabla w_s \;d\Omega \notag\\ &+\bar D\int_{\Omega}\bar \nabla \bar \rho \cdot \bar \nabla w_s \;d\Omega \;.
\end{align}

\subsection{Free surface finite element procedure}
The problem is formulated as a free-boundary problem: the computational domain is not prescribed a priori but is determined self-consistently from the deformation of the free surface. The objective is to deform the domain iteratively through several steps (labeled by $n$) until the free boundary $\partial\Omega_f$ coincides with the interface defined by $\bar{v}_z \bar{\rho} \approx 0$. The numerical procedure is as follows:
\begin{enumerate}[(i)]
\item The initial domain $\Omega_{n = 0}$ has a flat top surface located at $\bar{z} \approx 1$. 
This choice is convenient but not unique. Above (below) this height the vertical velocity component is negative (positive). The membrane boundary follows a fixed cosine-shape $u(x) = \cos(qx)$ independent of time.   
\item On the current domain $\Omega_n$, we solve the compressible Stokes problem as stated before. The resulting velocity field $\bar{\boldsymbol{v}}$ at the free surface is used to construct a domain deformation via harmonic extension \citep{shamanskiy2020mesh,donea2004arbitrary}. Specifically, we solve
\begin{alignat}{2}
\Delta \bar{\boldsymbol{v}}_{\textsf{\tiny harm}} &= 0\quad && \text{in } \Omega_n,\\
\bar{\boldsymbol{v}}_{\textsf{\tiny harm}} &= \bar{\boldsymbol{v}}\quad && \text{on } \partial\Omega_{f,n},\\
\bar{\boldsymbol{v}}_{\textsf{\tiny harm}} &= 0 \quad&& \text{on } \partial\Omega_{m,n}.
\end{alignat}
The harmonic field is then used to update the mesh by displacing each vertex $\bar{\mathbf{x}}_j$ according to $\Delta \bar{\mathbf{x}} = \bar{v}_{z,\textsf{\tiny harm}} \Delta t$, yielding the new domain $\Omega_{n+1}$. The lateral boundaries remain fixed in the $x$-direction, so only vertical displacements are applied. The pseudo–time step $\Delta t$ is chosen sufficiently small to avoid mesh distortion.
\item The procedure is repeated until convergence. The iteration is terminated once the normal flux measure defined as
\begin{equation}
    \mathcal{E} =  \max_{\partial \Omega_f} \left| \bar{\rho} (\boldsymbol{\bar v}\cdot \boldsymbol{n}) \right| \;,
\end{equation}
satisfies $\mathcal{E}$ $<10^{-6}$  approximating the location of the free surface. 
Note that the notion of a free surface is slightly different for the fully compressible case ($\bar{\chi} = 0$), for which the free surface is located at $\bar{z} \rightarrow \infty$ as the density is exponentially decaying (compare \cref{fully_compressible_limit})
\end{enumerate}

\subsection{Effect of diffusion}
As mentioned before, diffusion is added to stabilize the numerical scheme by damping out arising short-length scale oscillations. Based on the chosen scales the diffusion constant is $D = \bar{D} h_0 v_p$. In order for diffusion to dampen potentially occurring numerical oscillations at the mesh length scale $l_{\text{mesh}}$ but not to influence our results at physically relevant length scales $L_{\text{phys}}$, we impose  $l_{\text{mesh}}\lesssim D/v_p\ll L_{\text{phys}}$.
The two relevant physical length scales are the thickness of the actin layer $h_0$ and the membrane corrugation wavelength set by $1/q$. Hence, we obtain the following criterion 
\begin{equation}
    \bar{l}_{\text{mesh}} \lesssim \bar{D} \ll \text{min}(1,\frac{1}{\bar{q}})\;.
\end{equation}
Our choice of $\bar{D} = 10^{-2}$ fulfills this criterion with the smallest mesh grid size of $l^{\text{min}}_{\text{mesh}} = 5 \cdot 10^{-3}$ and $\bar{q}_{\text{min}} = 2\pi/10$ 

\subsection{The incompressible limit in the FEM}
In the large-$\chi$ limit, the equation of state $P=\chi(\rho-\rho^\ast)$ enforces $\rho\simeq\rho^\ast$, so the model approaches incompressibility. Numerically, this acts like a penalty formulation: increasing $\chi$ improves enforcement of the constraint but makes the recovered pressure sensitive to the finite accuracy of the density field. A residual density error $\delta\rho$, for example from interpolation or mesh-scale fluctuations, produces a pressure error $\delta P=\chi\,\delta\rho$, which can contaminate the computed stresses when $\chi$ is large.\\
Since pressures and stresses are nondimensionalized by $2\eta k_d$, and since the flat incompressible reference pressure is of order $P_s=-2\eta k_d$, the corresponding nondimensional smooth pressure scale is $|\bar P_s|\sim 1$. Therefore, a remaining density fluctuation $\delta\rho_m$ contaminates the computed stress field if $\bar\chi|\delta\rho_m|$ becomes comparable to unity. To estimate when such mesh-scale density fluctuations are suppressed, consider a perturbation $\delta\rho_m$ with characteristic wavelength $\ell_m$. In the density equation, diffusion contributes a residual of order $D\,\delta\rho_m/\ell_m^2$, and therefore penalizes short-wavelength fluctuations strongly. By contrast, the pressure perturbation $\delta P_m=\chi\,\delta\rho_m$ enters the compressible Stokes balance and induces a velocity perturbation, which re-enters the density equation through the transport term $\nabla\cdot(\rho\mathbf v)$. This pressure-flow feedback contributes to the density residual with a scale of order $\rho_s\chi\,\delta\rho_m/\eta$, where $\rho_s$ is the local smooth density. Hence diffusion robustly damps mesh-scale density fluctuations when $D/\ell_m^2 \gtrsim \rho_s\chi/\eta$. In nondimensional variables, this becomes $\bar\chi \lesssim \bar D/(\rho_s\bar\ell_m^2)$. This criterion should be understood as a mesh-scale damping estimate, not as a sharp physical transition. It states that a nonzero mesh-scale density fluctuation should generate a sufficiently large residual that the Newton solver cannot converge while retaining an appreciable $\delta\rho_m$.

With the values used in the simulations, $\bar D=10^{-2}$ and $\bar\ell_m\simeq 5\times10^{-3}$, this gives the estimate $\bar\chi_{\max}\sim 400/\rho_s$. Thus values such as $\bar\chi=5$ to $20$, which already approach the incompressible regime in our simulations, remain well below this mesh-scale damping bound. Nevertheless, the pressure-noise condition $\bar\chi|\delta\rho_m|\ll 1$ must also be checked, because even very small density errors may generate stress errors of physical magnitude when $\bar\chi$ is large. This distinction is important: diffusion controls the amplitude of short-wavelength density noise, whereas the equation of state determines how any remaining density noise is amplified into pressure and stress.
\begin{figure}[b]
     \centering
     \includegraphics[width=\linewidth]{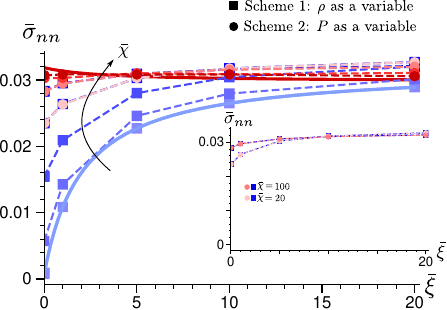}
     \caption{Numerical FEM results for the normal stress as a function of the friction parameter for the full range of compressibility $\chi$, using the two numerical schemes described in the text. In order to suppress residual density fluctuations in the large $\chi$ regime, the diffusion coefficient was increased to $\bar D=0.1$. 
     As the compressibility, $\bar \chi$ increases, the numerical solutions converge towards the bold red curve that corresponds to the analytical incompressible prediction including both first and second order contributions.  Parameters: $\chi=[0,1,5,20,100,500,800]$, $qh_0=\pi$.}
     \label{fig:incompressible}
 \end{figure}

To improve our scheme for larger values of $\bar \chi$ we reformulate the compressible problem by keeping the pressure as a variable and deriving the density from it:
\begin{equation}
    \rho = \frac{P}{\chi} + \rho^*\;. \label{p-reformulation}
\end{equation}
This strongly reduces the amplification of residual density errors for large $\chi$ and allows stable asymptotic numerical exploration of the incompressible limit as shown in Fig.\ref{fig:incompressible}. 
In the weak form, this reformulation modified the density in favor of the pressure variable as in \ref{p-reformulation} and the governing equations becomes
\begin{align}
F_P&=
\int_\Omega \bar{P}\, w_s \, d\Omega+\int_\Omega \bar{\chi}\rho^\ast w_s\,d\Omega-\int_\Omega\bar{P}\,\bar{\boldsymbol{v}}\cdot\bar{\nabla} w_s \, d\Omega\notag\\
&-\bar{\chi}\int_\Omega \rho^\ast\bar{\boldsymbol{v}}\cdot\bar{\nabla} w_s \,d\Omega+\bar{D}\int_\Omega\bar{\nabla}\bar{P}\cdot\bar{\nabla} w_s \, d\Omega\notag \\
&+\bar{\chi}\int_{\partial\Omega_f}(\rho^\ast\bar{\boldsymbol{v}}\cdot\boldsymbol{N})\, w_s \, ds+\int_{\partial\Omega_f}\!(\bar P  \boldsymbol{\bar v}-\bar{D}\bar{\nabla}\bar{P})\cdot\boldsymbol{N} w_s  ds.
\end{align}
We observe that increasing $\bar\chi$ drives the stress towards the analytical incompressible limit, including also the nonlinear corrections. This second scheme ($P$ as a variable) suffer from the same limitation for small $\chi$ as the first scheme ($\rho$ as a variable) for large $chi$, so a proper numerical exploration of the full range fo $\chi$ from zero to infinity needs to combine both schemes. \cref{fig:incompressible} shows such an exploration, insisting on the important requisite that both schemes give the same results for intermediate $\chi$ values.

\section{Additional results for finite compressibility.}\label{app:compressible}
\begin{figure}[t]
    \centering
    \includegraphics[width=\linewidth]{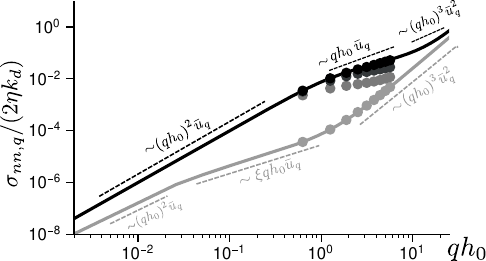}
    \caption{The curves includes the \textbf{nonlinear contributions} for the \textbf{incompressible and fully compressible} case. It is followed by the numerical results in the regime where the non linear effects become non-negligible for small friction value. Derivation of non linear terms: see Section \ref{non_linear_derivations}. Parameters used: $u_q / h_0 = 0.01$, $\xi h_0 / 2\eta = 0.01, \chi/(2 \eta k_d)=[0, 1,5,100]$. Nonlinear effects are expected to dominate when $\bar q^2>\bar\xi/\bar u_q\simeq 1$ (\cref{eq:non-linear})}
    \label{fig:scalings_nonlinear}
\end{figure}

\appendixfigures
Fig.\ref{fig:stressmap} shows how the stress from a gel with finite compressibility extrapolates smoothly between the fully compressible and incompressible solutions as $\bar\chi=\chi/(2\eta k_d)$ increases, for a given value of the wavevector $q$. When the friction coefficient is small, the fully compressible stress is dominated by second order contributions (\cref{eq:non-linear}). In order to properly compare the incompressible and fully compressible cases, we
performed the same second order expansion for the incompressible limit as described in \cref{non_linear_derivations} in the fully compressible case. The equivalent of \cref{eq:non-linear} for incompressible gel are very lengthy and not given explicitly here. It exhibits a similar scaling  $\propto q^3 u_q^2$. \cref{fig:scalings_nonlinear} display the analytical expressions for the incompressible and fully compressible stresses as a function of $q$, together with   the numerical simulations using FEM for varying compressibility. This shows that finite-compressibility effects become more apparent at large wavevectors $q h_0\gg1$.

 \clearpage


%


\end{document}